# Improved cooperation by balancing exploration and exploitation in intertemporal social dilemma tasks

**Cheng Zhenbo[a], Liu Xingguang[a], Zhang Leilei[a], Meng Hangcheng[b], Li Qin[a],**

**Xiao Gang[a]**

[a] College of Computer Science and Technology and College of Software, ZheJiang University of Technology, HangZhou, ZheJiang 310032, China
[b] College of Mechanical Engineering, ZheJiang University of Technology, HangZhou, ZheJiang 310032, China

Corresponding author.
E-mail address: czb@zjut.edu.cn (Cheng Zhenbo).

**Abstract:** When an individual's behavior has rational characteristics, this may lead to irrational collective actions for the group. A wide range of organisms from animals to humans often evolve the social attribute of cooperation to meet this challenge. Therefore, cooperation among individuals is of great significance for allowing social organisms to adapt to changes in the natural environment. Based on multi-agent reinforcement learning, we propose a new learning strategy for achieving coordination by incorporating a learning rate that can balance exploration and exploitation. We demonstrate that agents that use the simple strategy improve a relatively collective return in a decision task called the intertemporal social dilemma, where the conflict between the individual and the group is particularly sharp. We also explore the effects of the diversity of learning rates on the population of reinforcement learning agents and show that agents trained in heterogeneous populations develop particularly coordinated policies relative to those trained in homogeneous populations.

## 1 Introduction

The results of group behaviors of animals and humans are not only affected by the environment but also by individual behavioral strategies within the group. For example, the migration of animals and humans is affected by the environment in which they are located, and animals and humans tend to migrate to places rich in environmental resources. Simultaneously, the rewards that individuals in the group receive from the environment are also affected by the strategies of other individuals in the group. When the strategy of all individuals in the group is to migrate to resource-rich places, the rewards for individuals in that place gradually decrease. This type of individual rationality leads to group irrationality, which is the so-called intertemporal social dilemma (Hughes, Leibo, Phillips, Tuyls, & G Ra Epel, 2018). To obtain the optimal collective return in the intertemporal social dilemma, individuals in the group need to be able to weigh the individual short-term reward and the group long-term reward. However, it is not clear how to balance an individual's short-term reward with the group's long-term reward.

Multi-agent reinforcement learning (MARL) simulates the behavioral strategies of multiple agents in a dynamic and variable environment. Many researchers have found that MARL simulates how cooperation is formed in groups to obtain optimal solutions in various social dilemmas. These models often use intrinsic rewards to allow agents to form rational strategies for the group. Intrinsic rewards include aversion to inequality (Engelmann & Strobel, 2004), prosocial behavior (Aknin, Van de Vondervoort, & Hamlin, 2018), and enhanced reputation (Cuesta, Gracia-Lázaro, Ferrer, Moreno, & Sánchez, 2015). These intrinsic rewards inhibit the agent from adopting strategies that are detrimental to the group. For example, agents are assumed to have inequity-averse social preferences. They balance their selfish desire for individual harvests against a need to minimize the deviations between their own harvests and the harvests of others. To form intrinsic rewards, these models often assume that individuals in the group can directly obtain information from others, thus forming social preferences. However, it is difficult to provide a reasonable explanation for the formation of a cooperative mechanism for groups that can only indirectly obtain information about a few companions in the group, such as a fish colony and ant colony.

Based on the theory of marginal value (Charnov, 1976), in this study, we propose that agents can balance exploration and exploration by simply adjusting the learning rate, and then form cooperation in the social dilemma task to increase the total group income. In the reinforcement learning model, exploration is expressed as the agent choosing actions that cannot currently obtain the optimal reward to avoid the local optimal solution. Exploitation is the action that the agent chooses to obtain the best reward currently. To trade-off between exploration and exploration, we propose a learning rate based on deep Q-learning, which is defined as the difference between the individual's cumulative income and the target income. The agent can switch between exploration and exploration according to the learning rate to form a relatively optimal individual strategy for the group's total harvests.

## 2 Related research

How agents gradually form cooperative behaviors when they encounter social dilemmas has always been an important research issue in social sciences, economics, and psychology. By constructing a game in which two participants interact, Komorita and Parks et al. found that by setting the rewards of the two strategies of "worry" and "greed", two players can form cooperative behavior (Komorita & Parks, 1995). Using the evolutionary dynamics model, researchers found that tit-for-tat strategies (Axelrod & Hamilton, 1981), cooperating with people who directly help oneself (Nowak, 2006), or punishing others to obtain sufficient rewards (Gaechter & Fehr, 2002) can all promote the formation of cooperation. Although the above results provide possible factors that form group cooperation, they do not provide individual-specific strategies.

Because of the remarkable results of reinforcement learning in solving problems such as Go (Silver et al., 2017) and multiplayer cooperative games (Du et al., 2019; Jaques et al., 2019), many researchers have begun to use the MARL model to study the mechanism of how cooperation is formed in groups (Badjatiya, Sarkar, Sinha, Singh, & Krishnamurthy, 2020; Hughes et al., 2018). By setting the decision-making tasks of the agent and the strategic parameters required for the agent to form cooperation, researchers can use the MARL model to explain the possible strategic parameters of animals and humans when they form cooperative behaviors. Sequeira et al. proposed that agents create social attributes by exploring intrinsic motivations (Sequeira, Melo, Rui, & Paiva, 2011). Foerster et al. used the agent to model the learning results of other individuals to allow the agent to cooperate in multiple rounds of the prisoner's dilemma game problem (Foerster, Farquhar, Afouras, Nardelli, & Whiteson, 2018). Peysakhovich et al. found that when the agent pays more attention to the rewards of other individuals, the agent can form a prosocial strategy in Stag Hunt games (Peysakhovich & Lerer, 2017). Hughes et al. integrated the aversion to inequality into the internal rewards of agents so that when their own rewards are much greater than those of other individuals in the group, or when their own rewards are much less than those of other individuals in the group, agents adjust their strategies to form cooperation (Hughes et al., 2018). Jaques et al. transformed the influence of individual actions on the group into internal rewards to allow agents in social dilemma tasks to form cooperation (Jaques et al., 2019). Wang et al. proposed evolutionary deep reinforcement learning, and defined the past and future rewards of other individuals as the internal rewards of the agent to evolve cooperative strategies (Wang, Hughes, Fernando, Czarnecki, & Leibo, 2018). Khadka et al. designed a method to learn multiple strategies with shared replay buffers and dynamically select the best learners to gradually evolve the cooperation between multiple agents (Khadka, Majumdar, Nassar, Dwiel, & Tumer, 2019). Badjatiya et al. proposed designing a Status-Quo loss function to allow the agent to follow the status quo as much as possible, thereby evolving cooperative behavior in a social dilemma environment (Badjatiya et al., 2020). McKee et al. found that agents can obtain prosocial attributes by sampling their rewards from heterogeneous groups (McKee et al., 2020). Danassis et al. found that the agent can improve cooperative behavior by incorporating public signals (e.g., time, date, and other periodic numbers) in the learning process (Danassis, Erden, & Faltings, 2021). A common feature of the above models is that to form cooperative behavior, agents need to directly obtain the relevant information about all other agents. It is difficult for these models to provide a reasonable explanation for the formation of the cooperative mechanism for groups that can only indirectly obtain a small amount of peer information within the group, such as a fish swarm and ant colony.

Because of the dynamic changes and uncertainty of the environmental state, the agent either only uses existing experience for exploitation or takes the risk of not being able to improve reward by exploring, to improve its strategy. Therefore, exploration and exploitation have always been important research topics in reinforcement learning. When solving the problem of multi-armed gambling machines in the early days, epsilon-greedy (Leibo, Zambaldi, Lanctot, Marecki, & Graepel, 2017), upper confidence bound (Auer, 2002), and Boltzmann (Edman & Dhir, 2019) exploration were able to balance exploration and exploitation to obtain the best collective harvests. However, because of the sparse nature of the reward signal and the abnormal noise in the environmental state in the real environment, the above simple exploration strategies cannot improve the collective harvest.

A more general method is to design an internal reward function to form the internal motivation of the agent to guide the agent to explore through approaches such as curiosity (Zgonnikov & Lubashevsky, 2013). Curiosity includes discovering a new state or improving the accuracy of the agent's estimation of environmental changes (Leibo et al., 2017). This type of exploration strategy based on internal rewards may have a slow convergence speed, and a non-stationary exploration reward makes it difficult to form a fixed exploration strategy. As a result, memory-based exploration strategies and resampling Q-value exploration strategies have been developed to avoid the deficiencies of internal reward exploration strategies (Weng, 2020). However, the above single-agent exploration strategy is not necessarily suitable for multi-agent collaborative exploration.

In the case of multi-agent exploration, it is not only necessary to encourage agents to explore new states and overcome the problem of sparse reward signals but also to coordinate actions between agents to form cooperation to explore the environment. Agogino and Tumer defined a method for evaluating the effectiveness of the reward function of multiple agents in a small state space (Agogino & Tumer, 2008). Jaques et al. defined an intrinsic reward function for MARL, which encourages agents to take actions that have the greatest impact on the behavior of other agents, thereby obtaining collaborative exploration strategies (Jaques et al., 2019). Mahajan et al. introduced a mechanism for executing "commitment" exploration, which allows agents to explore a common strategy for temporary expansion (Mahajan, Rashid, Samvelyan, & Whiteson, 2019). Wang et al. defined influence-based rewards, which encourage agents to visit areas where their behavior influences the transformation and rewards of other agents (Wang et al., 2018). Recently, Iqbal and Sha proposed a type of exploration method based on internal rewards. The main feature of this method is that it can collaborate with the exploration strategy between agents and enable the agents to improve a collective return (Iqbal & Sha, 2018). The above multi-agent exploration strategies still needs agents to obtain other agents' information directly. We need to conduct further research on the exploration strategy under the condition that only a small amount of the other agents' information can be obtained, even if this information is not used.

## 3 Multi-agent reinforcement learning and decision-making tasks

## 3.1 Multi-agent reinforcement learning

We define the MARL model as a quad, which includes the state set $S$, state transfer function $T$, action set $A$, and reward $R$, that is, $< S, T, A, r >$. There are $N$ agents in the environment, and the state that each agent can perceive is $O_n: S \to \mathbb{R}^d$, which means that $agent_n$ can observe $d$ dimensions of the state; that is, the agent can only partially observe its state. Each agent in the environment interacts with the environment through its actions $A_n$, and the actions of the agent cause changes in the state of the environment. The change is described by the state transition function: $T = s \times A_1 \times A_2 \times ... \times A_n \longrightarrow s'$; that is, the actions of all agents in the environment work together to change the state of the environment from $s$ to another state $s'$.

Each agent $i$ learns the strategy $\pi(a_i|o_i)$ according to its observation $o_i = O(s,i)$. After the agent executes the action $a_i$, it obtains the reward $r_i$, and evaluates the result of the action through the reward. The goal of the agent is to learn an optimization strategy to obtain the greatest long-term reward. The definition of the long-term reward of an agent is

$$Q_{\vec{\pi}}(s_0, \overrightarrow{a_0}) = \mathbb{E}\left[ \sum_{t=0}^{\infty} \gamma^t \overrightarrow{r_t} | \overrightarrow{a_t} \sim \overrightarrow{\pi_t}, s_{t+1} \sim T(s_t, \overrightarrow{a_t}) \right],$$

(1)

where $\gamma$ is a discount factor between 0 and 1. For simplicity, $\vec{a} = (a_1, ..., a_n)$. For agent $i$ to obtain the maximum expected reward, the $Q$ function can be updated according to the following function (Watkins & Dayan, 1992):

$$Q_{t+1}^i(s_t, a_t) = Q_t^i(s_t, a_t) + \eta^i \left[ r_{t+1}^i + \gamma^i \max_{a'} Q_t^i \left( s_{t+1}^i, a' \right) - Q_t^i(s_t^i, a_t^i) \right].$$

(2)

## 3.2 Dynamical learning rate

We define the learning rate through the staged cumulative reward and target reward of the agent. The learning rate reflects the impact of changes in the environment on the agent's strategy. To achieve this goal, we define the difference between the stage cumulative reward $\tilde{R}$ and the target reward $\hat{R}$ as the learning rate:

$$\eta^i = max\{\hat{R}^i - \tilde{R}^i, 0\}/\hat{R}^i * \beta,$$

(3)

where $\beta$ is a constant, and the size is set to 0.001. The stage cumulative reward $\tilde{R}$ is the cumulative reward value of the agent in time span $\tau$, which reflects the indirect influence of other agents on the individual's reward in a certain period of time. The target reward $\hat{R}$ is a fixed value, and each agent has a target reward, which reflects the degree of satisfaction of the agent. When the target reward is large, this means that the agent needs to increase cumulative reward to be satisfied. If the agent's cumulative reward is less than the target reward, this indicates that the goal of the agent has not been reached, and it indicates that more exploration is required. When the agent's cumulative reward is close to the target reward, this indicates that the agent's strategy has reached its expectations, and it favors more exploitation.

According to the above definition of the learning rate, when the environment is in a stable state, the agent's strategy gradually converges, and its learning rate is at a low level. When there is a sudden change in the environmental state, the agent's strategy must be able to adapt to this change quickly, and the learning rate during this period is at a relatively high level.

### 3.3 Decision task

Following the resource collection task of (Hughes et al., 2018), we designed a similar intertemporal social dilemma task. The task environment contains two resource areas with different values: the apple area and garbage area. The size $S$ of the environmental map is $12 \times 20$ units, garbage is distributed in the upper half of the environment, and apples are distributed in the lower half of the environment. Garbage appears in its area with probability $\delta_g$, and the amount of garbage in the environment is recorded as $N_g$. Apples appear in their area with probability $\delta_a$, and the number of apples in the environment is recorded as $N_a$. The growth rate of apples is negatively correlated with the amount of garbage:

$$\delta_a = -\sigma/\Delta S_g * N_g + \sigma,$$

$$(4)$$

where $\sigma$ is the maximum growth rate of apples and $\Delta S_g$ is half the garbage area in the environment.

A number of agents are distributed in the environment, and agents move in the environment to either reap rewards or clean up the garbage at their location. The reward for the agent harvesting apples is recorded as $r_a$ and the reward for cleaning garbage is recorded as $r_g$. The purpose of the agent is to obtain the greatest collective return. In this task, each agent can only perceive the information in the limited field of view around it, and the field of view of the agent is denoted by $v$.

The dilemma of this decision task is as follows: the growth of apples and growth of junk affect each other. Because the reward for apples is greater than the reward for garbage, the agent's individual strategy tends to be to pick apples rather than clean up the garbage. However, a decrease in the number of apples leads to an increase in the amount of garbage, thereby suppressing the probability of apples appearing. Therefore, for the agent group, it is necessary for some agents to clean up the garbage and some agents to collect apples to obtain a more collective return overall.

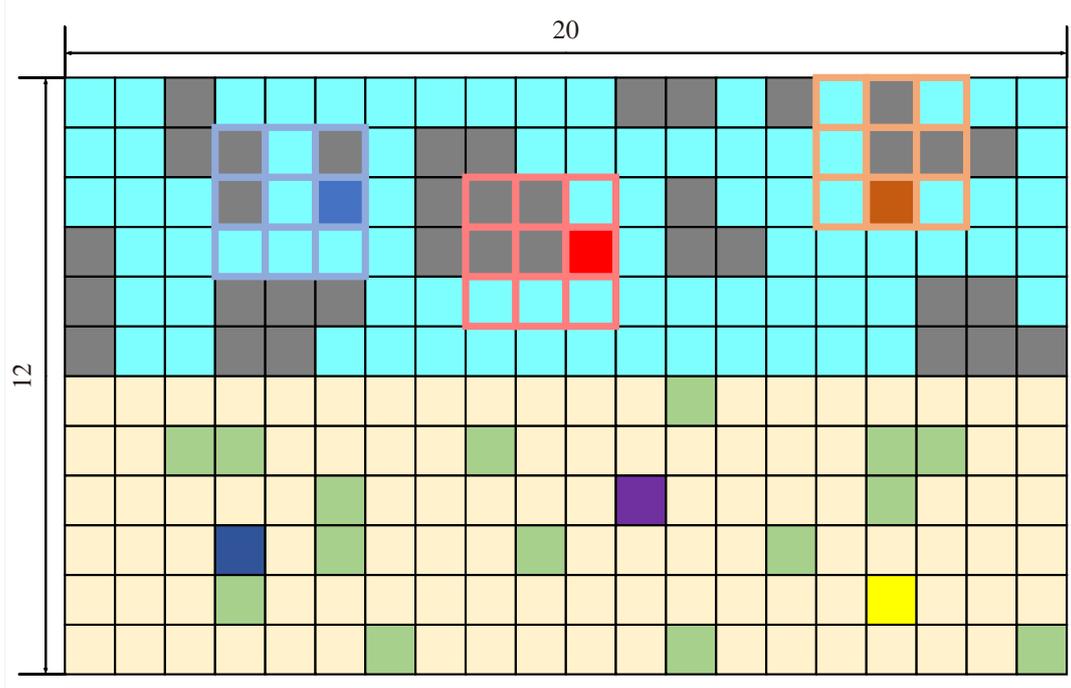

Figure 1. Game map

Because the growth of garbage and apples is not balanced, the multi-agent cleans up all the garbage in the field of vision when cleaning up the garbage $v_g$. When collecting apples, only the apples at the current location are collected. Therefore, the actual reward obtained by the agent for cleaning up garbage is $reward_g = v_g * r_g$. The actual reward for collecting apples is $reward_a = r_a$. We set $\tilde{R}^i = \sum_{t=1}^{\tau} reward_t$ in the decision task, which represents the collective return of the agent $i$ in time span $\tau$.

In the simulation, agents $Agent_{1-6}$ are randomly placed in the environment; their respective learning functions are shown in Equ. (1). We map the location of each cell in the environment to the interval [-120, 120], where [0, 120] represents the cell in the garbage area and [-1, -120] represents the cell in the apple area. $M_i$ represents the initial position of each agent in the environment, and $M_i \in [-120, 120]$. For each agent, each episode includes 100 trials, and each group of experiments includes 300 rounds of episodes.

### 3.4 Homogeneous and heterogeneous group attributes

According to the value method of $\hat{R}^i$, the agent group is divided into heterogeneous and homogeneous groups. Heterogeneity means that $\hat{R}^i$ takes a random value within a given range, and this heterogeneity reflects the diversity of individual agents. When the target reward of the agent satisfies $\hat{R}^i \geq reward_a * \tau$ or $\hat{R}^i \leq reward_g * \tau$, the agent is called as high target rewarder or low target rewarder. The details are presented in Table 1.

Table 1. Group attribute parameter settings

| Group attribute | | $\hat{R}^{i=1,2,3}$ | $\hat{R}^{i=4,5,6}$ | $M^i$ | $\tau$ | $r_a$ | $r_g$ |
|---|---|---|---|---|---|---|---|
| Heterogeneous | | $10 < \hat{R}^i$ | $\hat{R}^i > 50$ | $M^{i=1,2,3} \in [0,120]$ $M^{i=4,5,6} \in [-1,-120]$ | 5 | 10 | 5 |
| Homogeneous | high | $\hat{R}^i > 50$ | $\hat{R}^i > 50$ | | | | |
| | low | $10 < \hat{R}^i$ | $10 < \hat{R}^i$ | | | | |

## 4 Result

We use the stage cumulative reward $\bar{R}^i$ and the target reward $\hat{R}^i$ to define the dynamic learning rate $\eta^i$. We verified that the learning rate can form group cooperation by balancing exploration and exploitation in inter-period social dilemma tasks to obtain a relatively good collective return.

In Figure 2, we compare the collective return of groups performing inter-period social dilemma tasks for a fixed learning rate and dynamic learning rate. The group collective return for the dynamic learning rate converges to between 2200 and 2500. However, the collective return of a group with a fixed learning rate (= 0.001) only converges to between 1300 and 1600. The agent only uses random strategies, and the collective return converges to between 700 and 1000.

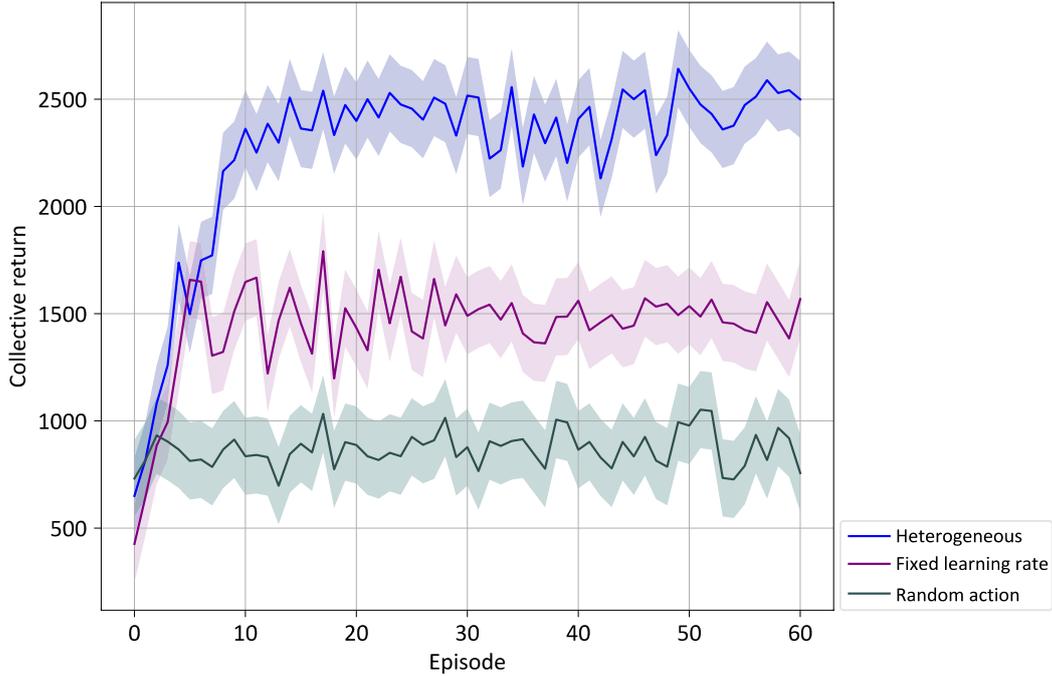

Figure 2. Comparison of the collective return between the dynamic learning rate and fixed learning rate

We set different distributions of target rewards to verify the role of the target reward in weighing exploration and exploitation. According to the distribution of the target reward, we defined three group attributes: heterogeneous, homogeneous high, and homogeneous low. Figure 3 shows that the collective return of the heterogeneous group is higher than that of the two homogeneous groups. The collective return of the homogeneous low group is the lowest; even lower than that of the random strategy.

To show the changes of exploration and exploitation of each agent in the environment more clearly, we drew the active position of each agent in the environment, as shown in Figure 4. heterogeneous groups conduct exploitation according to their own target reward to form a division of labor (Figure 4A). This type of division allows some agents (agents with low target rewards) to collect garbage in the garbage area and some agents (agents with high target rewards) to collect apples in the apple growth area. This distribution leads to the highest collective return for the heterogeneous group. When the target reward of each agent in the group is high, the agents have been using exploration in the environment to obtain high-yield apples (Figure 4B). Additionally, when the target reward of each agent in the group is low, the agents' reward in the apple area is greater than the individual's expectations, and the model inhibits low target rewarders from greedily collecting apples. They update their strategy when they explore the area that matches their own target reward (or junk area) (Figure 4C). The group that chooses actions completely at random has been performing goalless exploration (Figure 4D).

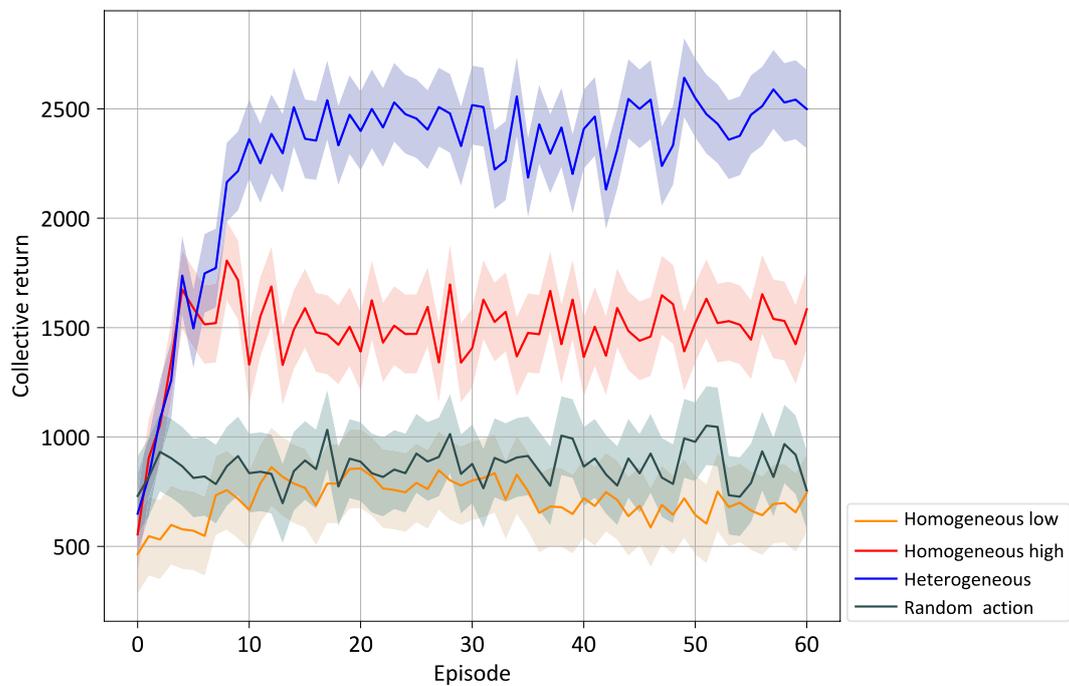

Figure 3. Comparison of the collective return between heterogeneous groups and homogeneous groups

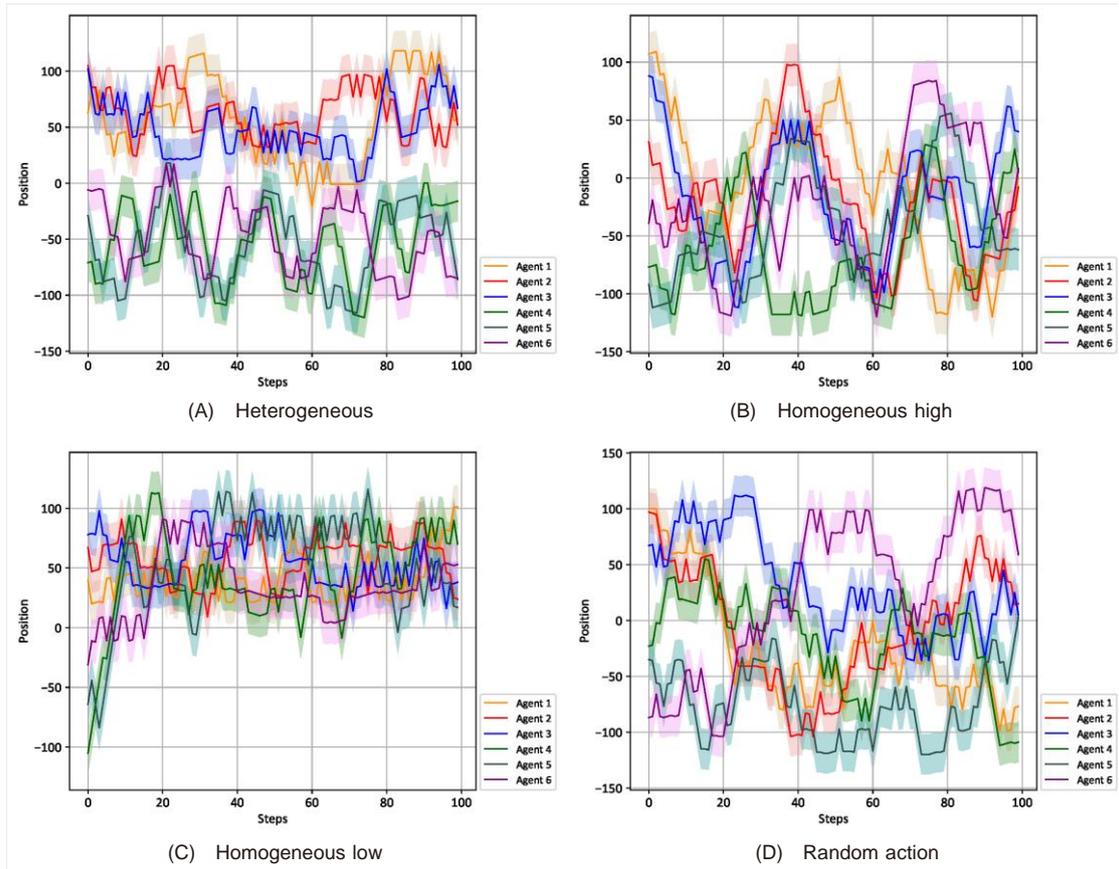

Figure 4. Comparison of the activity position of agents in a heterogeneous group and a homogeneous group

In Figure 5, the total reward between each agent in different groups is compared to determine whether cooperative behavior causes the gap between the rich and the poor. The results show that the reward difference of agents in the heterogeneous group is larger than the total reward difference of agents in the homogeneous group. This shows that although the heterogeneity of the group promotes cooperative behavior between agents, it also causes the gap between the rich and poor within the group to increase.

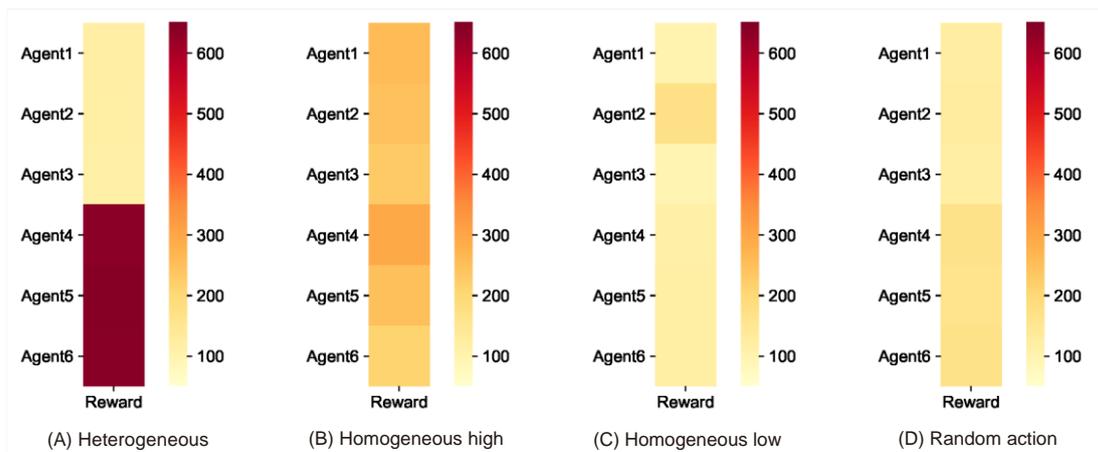

Figure 5. Comparison of the personal rewards of agents in different groups

We use the heterogeneous group as the experimental object separately, and control the target reward of each agent by leaving it unchanged to explore the influence of the cumulative length $\tau$ of different stages on group cooperation behavior. Figure 6 shows that as $\tau$ continuously increases, the collective return of the group continuously decreases; when the value of $\tau$ is small, the collective return of the agent is also small. When the value of $\tau$ is large, it is difficult for the agent to perceive changes in the environment, which reduces the possibility of it using exploration. When the value of $\tau$ is small, the agent only pays attention to the current reward, and it is also difficult to perceive changes in the environment, which makes the agent frequently adopt exploration.

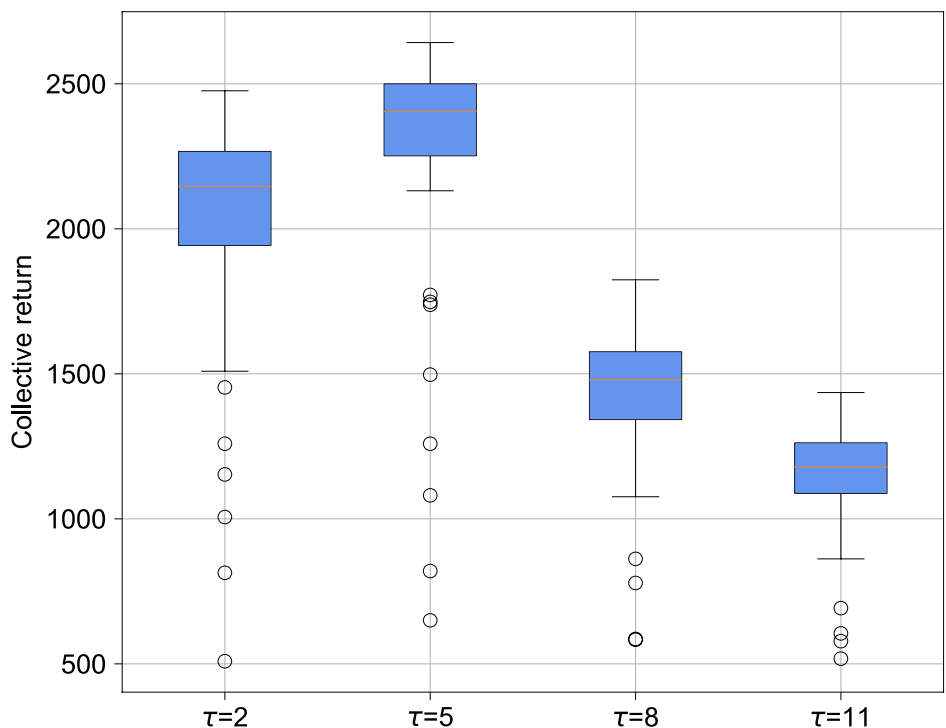

Figure 6. Collective return of the cumulative length $\tau$ in different stages

## 5 Discussion

The marginal value theorem proposes that people's desire for objects diminishes as they continue to be satisfied. Based on this theorem, we propose a method to form cooperation among multiple agents in intertemporal social dilemmas by balancing exploitation and exploration. The collective return of agents in intertemporal social dilemmas gradually converges through interaction with the environment, but the collective return may converge to a low level. Agents' strategies cooperate with each other to form a certain division of labor according to environmental changes and obtain a high level of overall collective return. The agent strategy algorithm we designed can coordinate the agent's respective strategies by balancing the agent's respective exploitation-exploration strategies according to changes in the environment. The parameters that affect the agent's strategy include cumulative rewards and target rewards. The cumulative reward is the reward the agent obtains from the environment in a period of time. The value not only reflects the benefits agents gain from the environment, but also reflects the impact of other agents on the environment. For example, when most agents gather apples in the apple area, the growth of new apples gradually decreases, and the cumulative reward of the agent also decreases. The target reward denotes the characteristics of the agents. If the value is large, the strategy of agents may include a large number of exploration features. Agent strategies are coordinated by agents adjusting the exploration and exploitation of their respective strategies. For example, when most agents are concentrated in the apple area to collect apples, the cumulative rewards of the agent gradually decreases, and then those agents with high target rewards increase the exploration in their strategy.

Our results show that the heterogeneity of target rewards among agents is the key to forming cooperative behavior. The heterogeneous performance of target rewards among agents allows agent groups to coordinate their strategies according to changes in the environment. The diversity of agent strategies is particularly important for cooperation in solving intertemporal social dilemmas. The problem of intertemporal social dilemmas requires agents to form a certain division of labor, which can ensure that their collective return is at a high level. This conclusion is similar to the recent research results of McAvoy et al. and McKee et al. (Mcavoy, Allen, & Nowak, 2020; McKee et al., 2020). They found that the heterogeneity of the number of connections of agents and the heterogeneity of social preferences can promote the formation of agents' social behaviors. Whether there are other heterogeneous parameters that promote the formation of cooperative behaviors of agents is worth following up in more in-depth theoretical and experimental research. Additionally, our results show that the cooperative behavior of agents increases the difference between the agents' rewards, which means that cooperative behavior increases the gap between the rich and the poor. This is because according to the strategy we designed, some agents with low target rewards always collect garbage in the garbage area, and the total reward of these agents is relatively low. How to make the agents cooperate and reduce the gap between the rich and the poor simultaneously is another issue worth further study.

One of the main shortcomings of our research is that we have only verified in the intertemporal social dilemma that the agent forms cooperation through adjustment, utilization, and exploration strategies. Whether the same conclusion can be drawn for other types of social dilemmas requires further verification.

## Acknowledgements

This work was partly supported by the National Natural Science Foundation of China (61976193). We thank Maxine Garcia, PhD, from Liwen Bianji (Edanz) (www.liwenbianji.cn/) for editing the English text of a draft of this manuscript.